# ARTICLE

# Entangled light-matter interactions and spectroscopy

Szilard Szoke*[a], Hanzhe Liu*[b], Bryce P. Hickam [b], Manni He[b], and Scott K. Cushing [b†]

Entangled photons exhibit non-classical light-matter interactions that create new opportunities in materials and molecular science. For example, in entangled two-photon absorption, the intensity-dependence scales linearly as if only one photon was present. The entangled two-photon absorption cross section approaches but does not match the one-photon absorption cross section. The entangled two photon cross section also does not follow classical two photon molecular design motifs. Questions such as these seed the rich but nascent field of entangled light-matter interactions. In this perspective, we use the experimental developments in entangled photon spectroscopy to outline the current status of the field. Now that the fundamental tools are outlined, it is time to start the exploration of how materials, molecules, and devices can control or utilize interactions with entangled photons.

## Introduction

Entanglement is arguably one of the most nonintuitive and fascinating phenomena in the quantum world. Entanglement refers to a many-body quantum state that cannot be decomposed into the product of each individual particle's state in the system[1]. Entangled states can be generated and measured in systems of particles including photons, electrons, and atoms. When a subset of particles in an entangled system interact with external stimuli, the many-body wavefunction of the whole system undergoes decoherence. This decoherence and the associated wavefunction collapse lie at the foundation of many technologies utilizing entanglement as a resource, such as quantum computation, communication, and information sciences[2].

In addition to these technologies, an emerging trend is to explore how entangled light-matter interactions differ from classical interactions. For example, entangled photons lead to two-photon absorption and sum frequency processes that scale linearly as if they are one-photon processes.[3–9] Using a ratio of the classical and entangled absorption rates, two-photon experiments should be possible at over a million times lower fluxes than classical experiments. Theoretical predictions suggest the same will be true for three and higher photon processes, converting nonlinear optical processes into linear processes.[10] Entangled photons have also been proposed for the excitation and control of excited state superpositions for qubits and molecular polaritons[11–13]. However, the origin of entangled light-matter interactions and the structural motifs in material and molecular design remain largely unexplored.

The exploration of entangled light-matter interactions currently resides in spectroscopic studies[14]. The central idea is to use the non-classical interference between entangled photon pairs to measure femtosecond and longer processes. Intriguingly, such measurements can be performed using only few-photon fluxes, allowing access to new intensity scalings and quantum phenomena[14]. Entangled photon spectroscopy also offers new possibilities when applied to more classical phenomena. Notably, entangled photons can break classical noise limits[15–19] and are predicted to break Fourier reciprocal spectral and temporal precisions.[20–23] For example, a 500 nm bandwidth of entangled biphoton pairs compressed to a few femtoseconds is predicted to only interact at the wavelength and linewidth specified by the input pump laser, usually <1 MHz for a modern Ti:Sapphire oscillator. A narrow excited state distribution with a high temporal resolution can therefore be created, potentially allowing new forms of quantum control[21].

At its heart, entangled photon spectroscopy relies on measuring changes in the quantum correlations of photons. The ability of entangled photons to couple, or not, with multi-particle excitations in materials is an intriguing question, both spectroscopically and from an application point of view. Entanglement has been proposed as fundamental to electron-electron and electron-phonon interactions in processes ranging from many-body correlations to singlet-triplet splitting[24–26]. In addition to the linearization of nonlinear interactions, we believe the potential for quantum correlated photons to create or interact with correlations in materials is why entangled-light matter interactions deserve further investigation in this field.

This article will provide a perspective on the current status of entangled light-matter interactions. Spectroscopic developments are used to outline the broader questions of how

[a.] Division of Engineering and Applied Sciences, California Institute of Technology, Pasadena, CA 91125, USA
[b.] Division of Chemistry and Chemical Engineering, California Institute of Technology, Pasadena, CA 91125, USA
* These authors contributed equally to this work
† scushing@caltech.edu; https://cushinglab.caltech.edu/

materials and molecules interact with entangled photons. Experimental exploration to date is limited to a few materials systems, partially due to the nascent development of these spectroscopic techniques. For example, entangled two-photon absorption has only been studied in a handful of molecular fluorophores.[6,27–33] However, the potential applications of entangled light–matter interactions make exploration and application in the materials field an intriguing possibility.

## Hong-Ou-Mandel interference & biphotons

To better understand entangled photon light-matter interactions, it is insightful to look at the fundamentals of the Hong-Ou-Mandel (HOM) interference.[34] The HOM effect serves as the basis for most entangled spectroscopy methods. A step-by-step guide to constructing an HOM interferometer can be found in reference [35]. In general, one pump photon is down-converted to create two lower energy daughter photons through spontaneous parametric down-conversion (SPDC). The daughter photons are collectively termed a biphoton. After temporal and polarization compensation in one arm, the two photons are made to meet and interfere at a 50:50 beamsplitter. Two single photon avalanche detectors (SPADS) then measure whether the photons leave opposite or same sides of the beamsplitter.

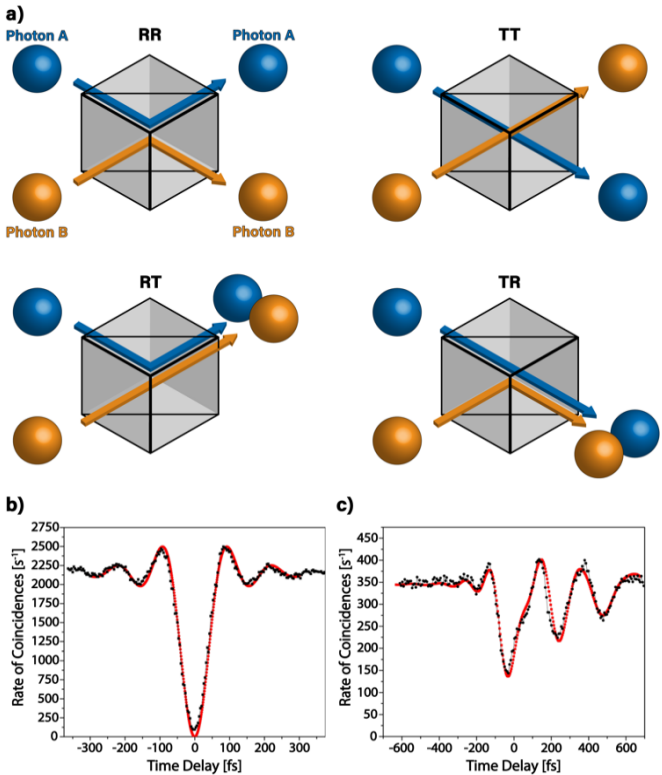

**Figure 1;** Hong-Ou-Mandel interference with photons. (a) Depiction of the four possible outcomes when two photons interact at a beamsplitter. When two entangled photons are present, the TT and RR outcomes cancel, and both photons leave the same side of the beam splitter. (b) This nonclassical interference leads to a dip in the coincidence counts. (c) HOM dip modulation due to a Nd:YAG crystal being placed in one arm of the HOM interferometer causing an entangled photon to interact with a sample with resonant states before the beam splitter. The excited state polarization is imprinted on the interference. Experimental results adapted from ref. 34.

Classically, there are four possible outcomes of transmission (T) and reflection (R) for the two photons, as depicted in Fig. 1(a): RR, TT, RT, TR. The indistinguishable character of the biphoton leads to the amplitudes of TT and RR destructively interfering in the entangled case[34], leaving the only measurable outcome of both photons leaving the same side of the beamsplitter. If the time delay or relative polarization between the photons is changed, or any other modification that partially distinguishes the biphoton pair, the interference will decrease. The width and amplitude of the interference dip therefore carries a signature of the light-matter interaction that the entangled photons have witnessed[36]. Measuring these variations in the interference dip enable the possibility for spectroscopy. As an example, Fig. 1(b) & 1(c) show the modulation of the interference dip when absorption is introduced in one arm of the HOM interferometer.[37]

For a full mathematical derivation of the HOM dip please refer to reference [38]. Briefly, the interference dip is explained mathematically as follows. Beamsplitter operators $\hat{B}_a$ and $\hat{B}_b$ are first defined which act on the bosonic mode operators to give a unitary transformation of an input state in ports a/b to the output in ports c/d.

$$\hat{a}^\dagger \xrightarrow{\hat{B}_a} \frac{1}{\sqrt{2}}(\hat{c}^\dagger + i\hat{d}^\dagger), \qquad \hat{b}^\dagger \xrightarrow{\hat{B}_b} \frac{1}{\sqrt{2}}(i\hat{c}^\dagger + \hat{d}^\dagger) \ (1)$$

$$\hat{B}_a |1\rangle_a = \hat{B}_a \hat{a}^\dagger |0\rangle_a = \frac{1}{\sqrt{2}}(|1\rangle_c |0\rangle_d + i|0\rangle_c |1\rangle_d) \ (2)$$

$$\hat{B}_b |1\rangle_b = \hat{B}_b \hat{b}^\dagger |0\rangle_b = \frac{1}{\sqrt{2}}(i|1\rangle_c |0\rangle_d + |0\rangle_c |1\rangle_d) \ (3)$$

With $a^\dagger$, $b^\dagger$, $c^\dagger$, and $d^\dagger$ being the creation operators for the photons in ports a, b, c, and d respectively, acting on the vacuum state $|0\rangle$. It is important to note that the beamsplitter operator introduces a π/2 phase shift in the reflected photon's output state. Next, assume a pair of entangled photons, each of which is defined by a singly occupied Fock state, is introduced into both input ports a and b:

$$\hat{B}_a \hat{B}_b |1\rangle_a |1\rangle_b = \frac{1}{2}(\hat{c}^\dagger + i\hat{d}^\dagger)(i\hat{c}^\dagger + \hat{d}^\dagger)|0\rangle_c |0\rangle_d \ldots$$

$$\ldots = \frac{1}{2}\left(i\hat{c}^{\dagger 2} + \hat{c}^\dagger \hat{d}^\dagger - \hat{d}^\dagger \hat{c}^\dagger + i\hat{d}^{\dagger 2}\right)|0\rangle_c |0\rangle_d \ (4)$$

The phase shift leads to the commutator term between the two photon modes in equation 4. Given that the photons are indistinguishable, the commutation relation for the creation operators equals zero,

$$[\hat{c}^\dagger, \hat{d}^\dagger] = [\hat{c}^\dagger \hat{d}^\dagger - \hat{d}^\dagger \hat{c}^\dagger] = 0 \ (5)$$

and the only terms left are the RT and TR terms such that both photons must leave the same side, resulting in a N00N state as the output:

$$\frac{1}{2}\left(i\hat{c}^{\dagger 2} + i\hat{d}^{\dagger 2}\right)|0\rangle_c |0\rangle_d = \frac{1}{\sqrt{2}}(i|2\rangle_c |0\rangle_d + |0\rangle_c |2\rangle_d) \ (6)$$

Consequently, only one of the two photon counting detectors will register a detection event. This results in the characteristic interference dip in the coincidence counting scheme (Fig. 1(b)). Entangled photon spectroscopy can therefore be thought of as a measure of how light-matter interactions modify the commutator in equation 5.

Before a discussion of the practical applications of SPDC, a few technical points should be clarified. First, HOM interference should be more correctly viewed as 'biphoton interference' since it relies on the indistinguishability of the two bosons as well as the underlying entanglement between them. Second, an HOM interference for a two-photon entangled state is only possible if the entangled biphoton wavepacket has a symmetric spectrum about the frequency degenerate diagonal, irrespective of whether the two photons are frequency degenerate or non-degenerate. Without this condition being met, the interference vanishes, and the beamsplitter becomes transparent with respect to the two input photons. In the case of two independent single-photon wavepackets, the condition for HOM interference indeed becomes that the two single-photon wavepackets must be identical. In practice, this means that care must be taken how the two-photon states to be used are generated and how the two independent beam paths are optically treated[39].

## Generation and detection of entangled photons

The two down-converted photons from SPDC display strong correlations in time, energy, and momentum due to the parametric mixing process. The time and energy correlations originate from the individual photons being generated simultaneously with energies that must sum to that of the pump photon. The momentum correlation is dictated by the phase matching condition. Energy-time entangled states can be viewed as the most general type of entangled states. Hyper-entangled photon states, e.g. entangled both in energy-time and in polarization, can lead to significant improvements in the experimental measurement statistics and the robustness of the spectroscopic setup.[40,41]

The simplest experimental approach to SPDC is using birefringent phase matching (BPM) in a $\chi^2$ nonlinear crystal such as β-barium borate or lithium triborate.[42,43] Generation rates of ~$10^3$-$10^4$ counts/s/mW are achieved with pump powers of a few tens of mW. Given that SPADs generally saturate at around $10^6$-$10^7$ counts/s, this generation rate is sufficient for coincidence counting experiments with most pulsed laser systems. However, due to the strict phase matching condition, the generated entangled photons have a narrow bandwidth. Other issues include cross-polarized pump/daughter photon combinations in Type-I and Type-II down-conversion, beam walk-off issues due to birefringence, a limited wavelength mixing range, and reliance on weaker nonlinear tensor elements.

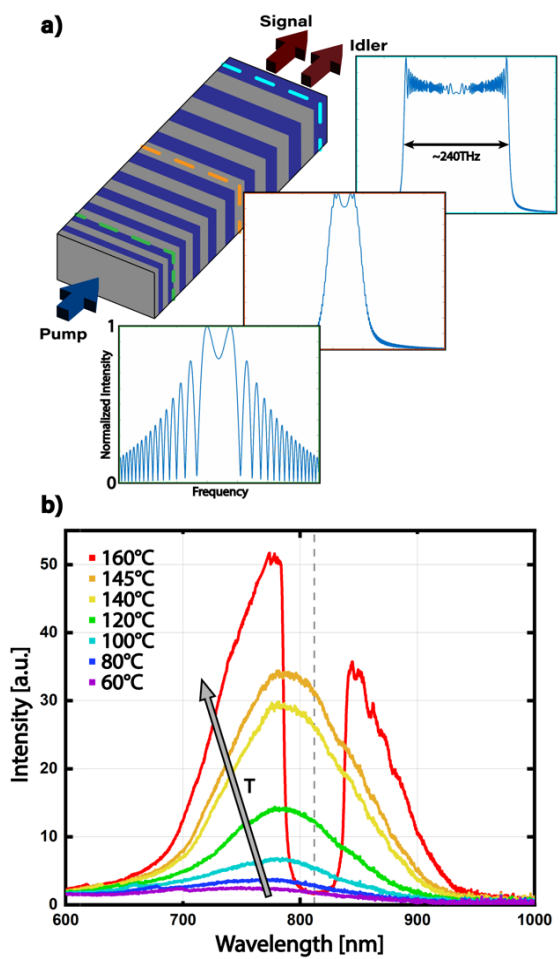

**Figure 2;** (a) Cartoon of a periodically poled waveguide with linear chirp, where purple and gray represent domains of opposing poling. Insets show the simulated downconversion spectrum at three points along the lithium tantalate waveguide; (green)-start, (orange)-center, (cyan)-end. The effect of the chirp is to broaden the output to a bandwidth of ~240THz (~510nm). (b) Experimental spectra of a periodically poled lithium tantalate grating where the working temperature of the crystal is tuned.

Quasi-phase matching (QPM) can generate broadband entangled photons more efficiently for spectroscopic applications. In this approach, a spatially periodic modulation of a ferroelectric nonlinear material is used to rectify the phase-mismatch of the three-wave mixing process, as illustrated in Fig. 2(a). This is achieved by using the additional momentum contribution from the crystal periodicity in the overall phase matching.[44,45] The QPM approach makes it possible to take advantage of stronger elements of the nonlinear tensor, as well as to implement Type-0 phase-matching whereby all three waves are co-polarized.[46] The down-converted photon flux can be as high as $10^9$ pairs/s/mW of the pump power by spatially confining the pump in chip-integrated photonic waveguides.[47] A particularly salient feature of QPM is that, since the crystal period can be arbitrarily chosen, it can be used to phase-match any desirable wavelength combinations. Further, by utilizing a longitudinally varying period, a collection of phase matching conditions can be used to create broadband SPDC fluxes.[48–51] Temperature controlled lithium niobate, lithium tantalate, and potassium titanyl phosphate are generally used given that they are transparent from the UV to mid-infrared wavelengths.[52,53]

Quasi-phase matching allows for two important advantages in spectroscopy. First, a sufficiently broadband flux of entangled photons can be used to increase average power levels without saturating the single photon per mode limit. As measured in our lab and others, almost microwatt fluxes of entangled photons spanning more than 500 nm can be created using QPM gratings.[54] The broad bandwidth allows for pulse-shaping and few-femtosecond resolutions. Second, the enhanced power levels allow spectrally and temporally resolved detection in reasonable time frames. In practice, this means that even microwatts of pump power would be enough to saturate a SPAD, suggesting that ultrafast entangled photon spectroscopy can be driven by a diode laser instead of expensive femtosecond laser amplifiers. Chip-integrated entangled photons are also easily fiber coupled. This provides alignment-free daily operations and an easy way to control and maintain the properties of entangled photons, such as spatial profile, over a broad bandwidth.

Multiplexed photon counting schemes are another area that could yield benefits in the application of entangled spectroscopy in materials science. Both EM-CCDs and SPAD arrays are coming to maturity, allowing spectral multiplexing of the photon counting process.[55–58] With a broadband, higher flux source, even a simple USB spectrometer can be used to measure spectral changes. Higher fluxes also allow phase-sensitive HOM techniques to be used which measure both 2nd and 3rd order correlations.[59] The difference between an HOM interferometer and a phase-unlocked HOM interferometer is conceptually similar to the difference between an autocorrelator and a frequency resolved optical gating (FROG) setup in ultrafast optics – the former only measures the intensity autocorrelation while the latter contains additional phase information.

## Spectroscopy with one entangled photon

One class of entangled photon spectroscopy utilizes the non-local nature of the entanglement. The general idea is that the interaction between the sample and one photon in the entangled pair can be revealed by measuring its entangled partner, even when it does not interact with the sample directly. Experimentally, one-photon interaction can be introduced by placing a sample in one arm of the HOM interferometer shown in Fig. 3. The sample imparts phase and amplitude changes on the entangled photon it interacts with, which affects the biphoton indistinguishability and modulates the HOM dip[37].

While it seems trivial to measure absorption profiles with entangled photon interference, the technique allows spectroscopy and microscopy at one wavelength using vastly different wavelengths[60–62]. Spectroscopic signatures, such as absorption spectra are reconstructed by monitoring the changes to the coincidence counts when a sample is inserted into the signal arm of the HOM interferometer and resolving the wavelength of the idler arm.[63–65] For example, if the SPDC process creates an entangled IR and visible photon pair, the absorption of the photon at the IR wavelength can be inferred by measuring its visible partner.[66] This particular case has been demonstrated in $CO_2$ as shown in Figure 3.[60] This effect is predicted to work for any range of experiments, such as X-Ray[67,68], THz, and electron spectroscopies, although practical limitations arise from generating such ultra-broadband SPDC sources.

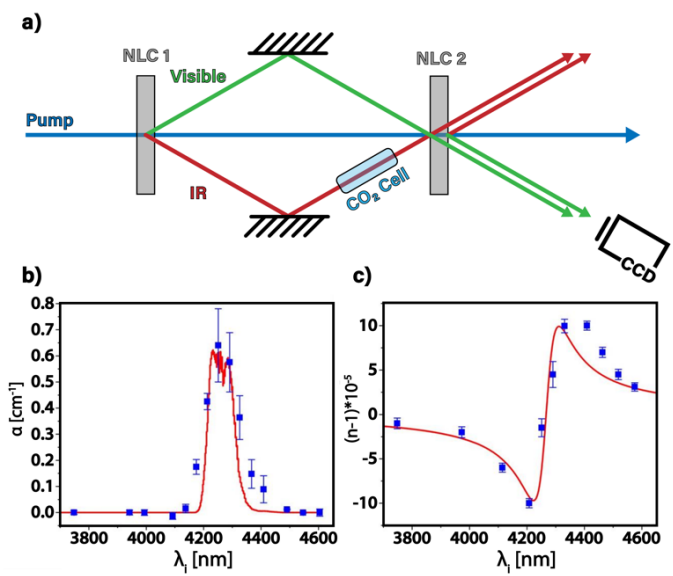

Figure 3; Measuring infrared absorption with visible photons. (a), Experimental layout. Entangled infrared (IR) and visible photons are generated through SPDC on the first nonlinear crystal (NLC1). A $CO_2$ cell is placed in the IR path to introduced absorption. The IR and visible photons cross at the second nonlinear crystal (NLC2) and interfere with the SPDC photons from NLC2. The interference pattern for the visible photons, which do not interact with the $CO_2$ are recorded. (b), IR absorption spectrum extracted from visible interference measurement. (c) refractive index near the $CO_2$ resonance. Adapted from ref. 57.

Entangled one photon spectroscopy has been applied to applications ranging from remote sensing to ghost imaging. For example, quantum ghost imaging relies on the coherence between the down-converted beams to record the image of an object with photons that do not interact with it directly.[62,69–75] Quantum optical coherence tomography (QOCT) offers improved resolution and sensitivity by exploiting the dispersion cancelling properties of the entangled wavepacket, as well as the anticorrelation between entangled pairs, to construct quantum interference patterns corresponding to sample depth.[76–78] Entangled photons can also increase measurement sensitivities as compared with classical photons. The standard quantum limit for noise scales as $1/\sqrt{N}$, where $N$ is the number of measurements. Using entangled photons, this improves to $1/N$.[79] A prototypical example is using N00N states to enhance the measurement precision of the phase shift in an interferometer.[80,81] Sub-shot-noise imaging of weak absorbing objects has also been achieved.[17]

## Spectroscopy with two or more entangled photons

Multiple entangled photon interactions can be split into two categories: the linearization of nonlinear interactions by preserving the correlations between the entangled photons (i.e. entangled two photon absorption), and measurements in which a sample modifies the correlations between photons to measure excited state properties. Multiphoton entangled experiments are performed by replacing the beam splitter in an HOM interferometer so that both entangled photons interact with a sample. Whereas many one-photon entangled interactions can be reproduced with shaped classical light, the interactions of two or more entangled photons with a sample lead to non-classical processes[14,23,82]. These interactions arise directly from the two-photon indistinguishability. Overly generalized, when the sample interacts with the biphoton pair, its response is as if only one photon is incident with the sum of their energy. Before being spatially and temporally overlapped, the two entangled photons propagate in the material as if separate photons.

The two-photon interaction can therefore be modulated by time-delaying or shaping one side of the HOM interferometer. This allows for the measurement of excited state polarizations as well as populations. For example, multidimensional spectroscopy can be recreated by using three or more entangled photons or pulse shaping the entangled photons.[14,83,84] Compared to conventional multidimensional spectroscopy, entangled multidimensional spectroscopy is predicted to suppress the uncorrelated background levels and enhance the sensitivity to electronic couplings, manifested as pronounced off-diagonal cross-peaks in the 2D spectra, Fig. 4.

There are other distinct differences between nonlinear entangled and classical spectroscopy. First, the two entangled photons are predicted to act as one only if their coherence / entanglement time, represented by the width of the HOM dip, is shorter than the decoherence of the excited state being measured. The coherence / entanglement time can therefore be used to alter and control the nonlinear process, allowing another route to multidimensional spectroscopy.[85] The measured spectrum is again more sensitive to electronic coupling than in the classical case. Second, temporal dynamics in entangled photon experiments are measured via the correlations between the two entangled photons. Unlike pump-probe spectroscopy, the temporal dynamics are not inferred from the sample's impulse response to a multiphoton pump pulse. Third, the entangled photons can measure whether an excited state superposition they excite in a qubit preserves their entanglement. Entangled photon interactions are therefore predicted to be sensitive probes for many-body dynamics and collective states[24,86,87].

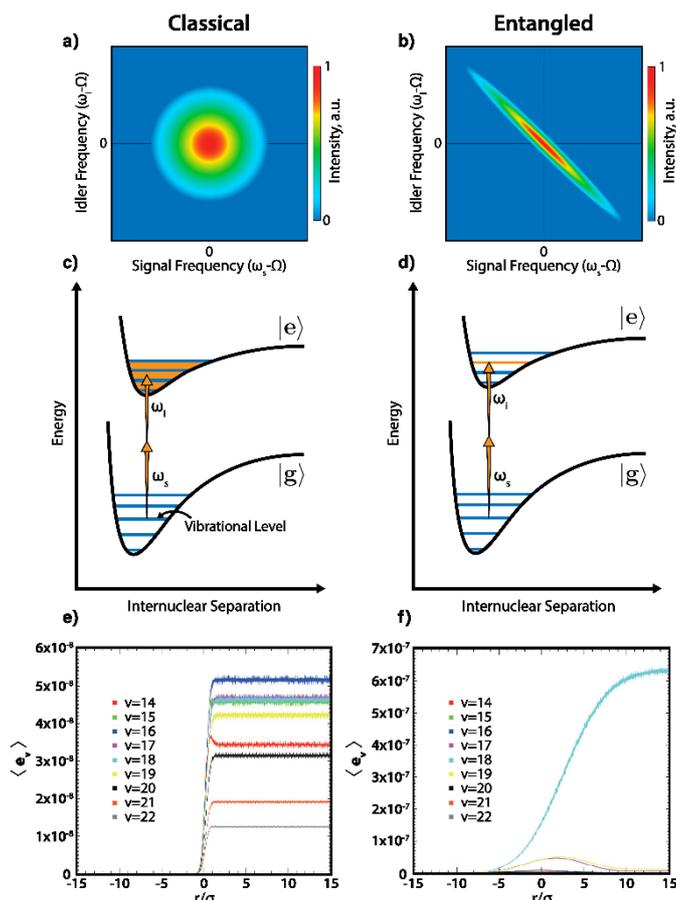

Figure 5; Spectral resolution in entangled two-photon spectroscopy. (a), Gaussian frequency distribution for uncorrelated photons as common from a classical laser. (b), frequency anti-correlation for entangled photon pairs from SPDC. (c), The classical frequency distribution means that any combination of photons can excite the sample. A pulsed laser therefore excites an ensemble of vibrational states as predicted in (e). (d), The entangled photon frequency distribution means that transitions only occur that add up to the pump source's linewidth. Selective excitation of a single vibronic level therefore is predicted to occur independent of the SPDC bandwidth and temporal resolution (f). (e), excited states population with pulsed laser excitation, where multiple vibronic levels are populated. Theoretical data (e) and (f) adapted from ref. 21.

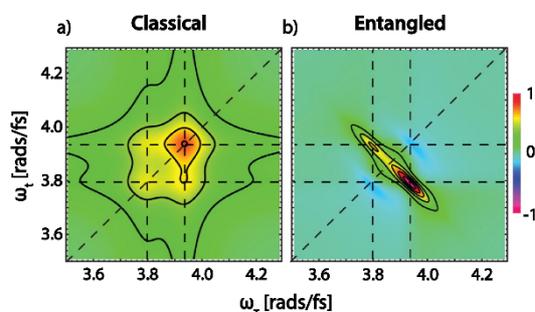

Figure 4; Simulated two-dimensional fluorescence spectra for electronically coupled molecular dimer with (a), classical light sources, and (b), entangled photon pairs. Multidimensional spectroscopy with entangled photons has improved sensitivity to the dimer conformation. Adapted from ref. 80.

Entangled photon spectroscopy also has the potential to measure ultrafast dynamics with higher spectral resolution than a classical approach. For entangled photon spectroscopy, the energy resolution is given by the down-converted pump source's linewidth. This is because the frequency-frequency photon distribution created by SPDC is correlated to the down-converted center frequency as shown in Fig. 5.[21] This is compared with the Gaussian distribution commonly associated with a laser, in which any two photons can interact with the sample (Fig. 5(a) compared to Fig. 5(b)). In an entangled interaction, only the frequency conjugate pair will interact with the material, post-selecting the photons along the frequency diagonal. All entangled photon interactions therefore add up to within the pump linewidth of the original pump laser source for the SPDC, and the frequency resolution is decoupled from the temporal resolution (Fig 5(c) compared to 5(d)). Broadband entangled two-photon absorption in rubidium has already demonstrated a 3 to 5 orders of magnitude improvement on spectral and temporal resolution.[88] It has also been suggested that the creation of only conjugate electric fields optimizes the otherwise nonlinear interaction.

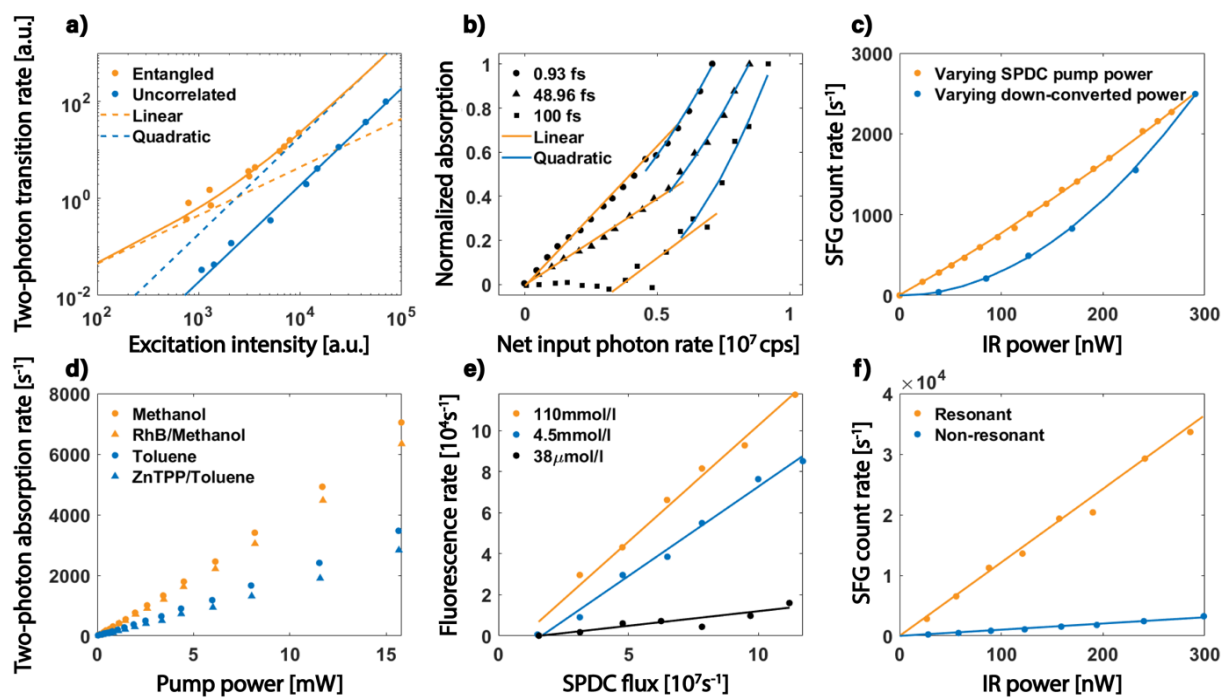

**Figure 6;** The linearity of various entangled two-photon processes as a function of power. (a), two-photon transition rate in trapped cesium with entangled and coherent light. Transition rate for uncorrelated coherent excitation is reduced by a factor of 10 for comparison. (b), power dependence of ETPA rate in porphyrin dendrimer under 3 different entanglement times. (c), power dependence of sum-frequency generation with entangled photons. (d), ETPA rate for RhB and ZnTPP in solvent. (e), ETPA-induced fluorescence rate for Rh6G in ethanol under different concentration. (f), resonantly enhanced sum-frequency generation with entangled photons. Data presented are adapted from Ref. 7, 6, 4, 30, 29 and 9.

The temporal resolution of the entangled experiment is still proportional to the spectral width the bandwidth of the down-converted photons. The width of the HOM dip can therefore be changed by modulating the relative time delay of the different frequencies using a pulsed shaper[88]. Combining pulse-shaping and entangled photons is predicted to allow non-classical population distribution and novel photochemistry processes.[88–91] The decoupled temporal and spectral resolution has been predicted to create narrow bandwidth excited state populations from femtosecond bandwidths as shown in Fig. 5(e) compared to Fig. 5(f). Simulation also suggests that under entangled two-photon excitation, the intermediate single-exciton transport in the bacterial reaction center of Blastocholoris viridis can be suppressed and non-classical control of two-exciton states can be achieved.[89]

## Current directions in entangled light-matter interactions

Several intriguing directions exist in exploring entangled light-matter interactions. One entangled photon interaction suggests that material and molecular interactions can modulate the probability of measuring a second, non-interacting photon. The role this could have in transmuting material and chemical changes, as well as in processes like energy transfer, leave many open questions. Perhaps most intriguing are interactions with two or more entangled photon interactions. Oversimplified, when two entangled photons are incident on a sample, they can appear as one photon with their summed energy. Nonlinear, multiphoton interactions up to N photons are therefore predicted to scale linearly.[10] The linearity of two photon entangled process was previously measured in sodium and cesium[7,8] and has now been repeated in some molecular and solid-state systems (Fig. 6).[4,6,7,9,32,33] Similar to the beamsplitter in the HOM interferometer, the strong amplitude and phase correlation[92] within the entangled photon pair of SPDC suppresses higher order interaction terms[88].

While the entangled two-photon intensity scaling can be understood, the cross section of the process raises many questions. The entangled photon cross sections measured to date in molecular systems fall within the $10^{-17}$-$10^{-18}$ cm$^2$ range[6]. This cross section is closer to that of a single photon absorption event, $10^{-16}$ cm$^2$, and orders of magnitude larger than that of a classical two-photon absorption event, $10^{-47}$ cm$^4$s. Why the entangled two photon absorption cross section does not match the classical one photon cross section, despite both processes being linear, is an open question. The entangled two-photon absorption cross section also does not appear to follow classical two-photon absorption rules in organic porphyrin dendrimers.[6]

In practice, the entangled two photon absorption will compete with the semiclassical two photon absorption. Their relative contribution is quantified by the ratio between the entangled absorption rate $\alpha I$ and the uncorrelated two photon absorption rate $\beta I^2$, where $\alpha$ and $\beta$ are associated cross sections. An entangled process is more efficient ($I < \alpha/\beta$) for fluxes lower than a GW for currently measured cross sections. The ratio can also be used to see that, for example, a one $\mu$W entangled photon flux has the same excitation rate as a million times more powerful classical laser. When calculating these ratios, it is important to include the conversion from W to photons/s through the ratio of 1 J = 6.242x10$^{18}$ eV for the $\beta I^2$. This must be done to cancel the cm$^4$s units when going to photons/s.

To study ultrafast material dynamics, an important factor in entangled two photon absorption is the coherence (entanglement time) relative to the decoherence of the excited state. Studies on several diatomic molecules and organic porphyrin dendrimers have suggested that transitions involving virtual states respond to entangled two-photon excitations at a different timescale than transitions involving other mechanisms, such as excited state charge transfer.[6,93] The entangled two-photon absorption cross section may therefore vary drastically between different transitions, exhibiting nonmonotonic dependence on the entanglement time, even if two systems classical cross sections are similar.

However, few experimental systems have been measured to date to confirm these rules. Measurements have yet to be repeated in condensed matter, low dimensional materials, or other common molecules – leaving many open questions as to how two or more photon entangled interactions can be optimized.  The primary questions can be summarized as: 1) How do intermediate states control the entangled photon interaction, 2) what structural motifs can increase or decrease the strength of the entangled photon interaction, 3), how does excited state coupling with spins, vibrations, and electrons preserve or decrease the entangled light-matter interactions, and 4), more generally, how to utilize entanglement to reveal and engineer novel materials and device responses that are not accessible with traditional methods. Other open questions, such as how selection rules are modified by polarization entangled photons and how photonic enhancement techniques will modify the interactions[94–102], promise for intriguing expansions to existing fields.

Controlling the entangled multiphoton interactions has practical as well as fundamental motivation. Entangled two-photon processes can occur at the same rate as a pulsed-laser-induced two-photon excitations but at over a million times lower fluxes, accounting for the intensity scaling between quadratic and linear processes. If the material and molecular design parameters can be optimized, optoelectronic and biomedical applications using two photon processes could be driven by a CW laser diode instead of a pulsed laser. Theory has also shown that due to the linear scaling, the spectrally overlapped simulated Raman scattering and two-photon absorption can be separated and selectively excited by tuning the entanglement properties of the pump.[103] For imaging applications, entangled two-photon fluorescence is suitable for in-depth imaging of photosensitive tissues at low flux. The linear scaling and enhanced cross section are predicted to occur for $N$-photon interactions, potentially bringing linearization to a family of nonlinear optical techniques.[10] Entangled multiphoton interactions have also been predicted to increase resonance energy transfer by several orders of magnitude[104] – and it can be extrapolated that the entangled effects will extend to most applications of multiphoton processes.

## New opportunities in materials science

The key distinction of entangled photon excitation, aside from the linearization of nonlinear processes, is that the entangled photons can be made to show quantum correlations in multiple variables (polarization, energy, OAM, etc)[105]. By interfering these entangled states of light with the quantum correlated excitations of a material, new properties and applications are to be expected[106]. Or, it may be possible to dynamically form correlations between two independent excitations that are not regularly present[107–111]. This key difference may lead to new insights in spectroscopy, but perhaps more excitingly, new degrees of control in quantum materials.  In this section, we outline a few potential directions that entangled-light matter interactions could have an immediate benefit in.

First and foremost, the study and understanding of entangled light matter interactions could become critical in the age of quantum computing and information systems[112,113]. How the entangled photon correlations are modified, favorably or not, by material elements is key for coupling qubit systems together as well as for investigating the types of control schemes that could be applicable to complex quantum systems[114]. Perhaps even more fundamentally, entanglement is suggested to be at the heart of quantum materials, which display strong electronic and nucleon correlations[115]. The inclusion of correlations in any Hamiltonian is the key to describing complex phenomena and for accurate calculations[116]. Many such Hamiltonians are also proposed to include entanglement, such as singlet-triplet scattering[25]. While the Fermi liquid theory and the Hubbard model have been used with great success, the direct experimental measurement of these strong interactions remains a difficult challenge[117–119]. The change in entangled photon correlations interacting with such systems may prove key, complementing approaches like ultracold atoms in optical lattices[120] and second quanta scans in multidimensional spectroscopy[121].

The ability of entangled photons to perform nonlinear optics with low power CW lasers should also prove transformative for photonics and electro-optical devices. Common components such as frequency mixers, saturable absorbers, or phase shifters can operate without a pulsed laser input, while wide band gap materials can be interacted with via photons at half the band gap energy. Additional functionality can be implemented in such devices relying on the preservation or change of the entangled photon correlations. An experiment that demonstrates this concept is to create polarization entangled biphoton states and couple one of the photons into a plasmonic metamaterial[122]. The coupled polarization state is determined and modulated by detecting the polarization of its partner. By altering the coupled polarization non-locally, the plasmonic device could in principle operate from perfect absorption to full transmission. The general idea of manipulating the material's functions non-locally via entanglement could lead to multifunctional integrated devices such as quantum logic gates, but also has fundamental advantages in processes like resonant energy transfer[104,123].

An important goal in materials science is the on-demand manipulation of electrons in solids via the application of external stimuli[124]. The coupling of the entangled photon pair's quantum correlations to strongly correlated materials could result in novel, exotic responses. One example would be in the field of valleytronics[125,126]. In two-dimensional semiconductors such as $MoS_2$, distinct valleys can be populated under a resonant excitation by circularly polarized light[125]. The circularly polarized excitation photon can be replaced by an entangled photon pair. More precisely, a cross-polarized pair could be generated via type-II down-conversion, after which the biphoton is split into its orthogonal polarization modes via a polarizing beamsplitter. A subsequent change in basis to circular polarization can be achieved by a pair of quarter-waveplates in both arms of the HOM type setup. The entangled photon energies could be set to select one or both valley excitations. The superposition between the spin state and the entangled photons would allow measurements of spin decoherence and coupling between the valleys. A similar idea could be used to test the ability of topological insulators to maintain spin correlations relative to photonic correlations.

Entangled photons could also find applications in material fabrication and lithography. Proposed techniques such as quantum photolithography aim to utilize entangled photons prepared in a N00N state to overcome the diffraction limit by a factor of $1/N$, alleviating the need to go to shorter UV wavelengths[127]. The enhanced cross-section for entangled multiphoton absorption could allow simplified photolithography and growth schemes, techniques that use below band gap light for patterning, or even additional degrees of freedom by using two or more entangled states. The microscopic chemical processes that govern photopolymerization or photocatalytic growth processes might similarly be tuned by entanglement.

## Outlook

While much work has gone into exploring entangled photons for quantum information and computational applications, the basics of entangled light-matter interactions and the potential applications in chemistry and material science are largely unexplored. In general, the field is advancing rapidly on the theoretical front, but most experimental questions remain open. How materials and molecular design can control entangled light-matter interactions remains mostly unknown, especially in materials and condensed matter systems. Given the relative ease of high-flux entangled photon generation, the simplicity of the optical schemes, and the various potential advantages of using non-classical light, we foresee entangled light-matter interactions being a rapidly growing field in the near future.

## Conflicts of interest

There are no conflicts to declare.

## Acknowledgements

This material is based upon work supported by the U.S. Department of Energy, Office of Science, Office of Basic Energy Sciences, under Award Number DE-SC0020151.

## Notes and references